\documentclass[twocolumn,showpacs,aps,prb,amsmath,amssymb]{revtex4}

\usepackage{graphicx}

\begin{document}

\title{Elementary excitations and the phase transition in the bimodal Ising spin 
glass model}

\author{N. Jinuntuya}
\affiliation{Department of Physics, Faculty of Science,
Mahidol University, Rama 6 Road, Bangkok 10400, Thailand}

\author{J. Poulter}
\affiliation{Department of Mathematics, Faculty of Science,
Mahidol University, Rama 6 Road, Bangkok 10400, Thailand}

\date{\today}

\begin{abstract}
We show how the nature of the the phase transition in the two-dimensional bimodal Ising spin glass
model can be understood in terms of elementary excitations. Although the energy gap with the ground
state is expected to be $4J$ in the ferromagnetic phase, a gap $2J$ is in fact found if the finite
lattice is wound around a cylinder of odd circumference $L$. This $2J$ gap is 
really a finite size effect that should not occur in the thermodynamic limit of the ferromagnet. 
The spatial influence of the frustration must be limited and not wrap around the system 
if $L$ is large enough. In essence, the absence of $2J$ excitations defines the ferromagnetic 
phase without recourse to calculating magnetisation or investigating the system response to
domain wall defects. This study directly investigates the response to temperature. 
We also estimate the defect concentration where the phase transition to
the spin glass glass state occurs. The value $p_c = 0.1045(11)$ is in reasonable agreement
with the literature.
\end{abstract}


\maketitle

\section{Introduction}
Spin glasses \cite{BY86,MPV87,FH91,KH04} have attracted much interest for quite a while.
Due to the considerable complexity of real materials much computational effort has gone into 
studies of a simplified model \cite{EA75} that is nevertheless thought to include the essential 
ingredients that lead to spin glass behaviour. Since even this model is not trivial, a
considerable industry has developed over time devoted to particular models
that, although probably unphysical, have provided subjects for the development
of numerical techniques \cite{HR01,Hart09}.

Systems known as spin glasses are disordered magnetic systems characterised by a 
random mixture of ferromagnetic 
and antiferromagnetic exchange interactions leading to frustration \cite{Toulouse}.
Typically, at low temperatures below a critical temperature $T_c$, a system undergoes a 
phase transition from a ferromagnet to a spin glass at some critical concentration $p_c$ of 
antiferromagnetic interactions. 

The model studied in this work is the bimodal, or $\pm J$,
Ising spin glass in two dimensions. This system has quenched bond (short range, nearest neighbour) 
interactions of fixed magnitude $J$ but random sign. The concentration $p$ of negative, 
or antiferromagnetic, bonds is varied from zero up the canonical spin glass at $p=0.5$.
It is believed that the spin glass can only exist at zero temperature \cite{ON09} where
$p>p_c$ with \cite{WHP03,AH04} $p_c=0.103$. This is clearly below the concentration
$p_n=0.109$ at the (finite temperature) Nishimori point, indicating a reentrant phase transition
as confirmed by Monte Carlo work \cite{TPV09}. 

The ground state is highly degenerate with
an entropy per spin \cite {BGP98,LGM04} of $0.07k$. Consequently spin correlation functions
are not guaranteed to take values $\pm 1.0$. If a nearest-neighbour bond correlation function 
does have a value $\pm 1.0$ then we call that bond a rigid bond \cite{BMR82}.
This means that the spin alignment across the bond is the same in all ground state configurations.
A recent study \cite{RRNRV10} suggests that the rigid lattice does not percolate in the spin
glass phase. This is consistent with the idea that the ferromagnetic phase is
characterised by percolation of rigid bonds.

Droplet theory \cite{McMillan84,BF86,FH86,BM87,FH88} has enjoyed much success with regard to
understanding the spin glass phase. The essential idea is that reversing all the spins
in a compact cluster with respect to a ground state provides a low energy excitation. 
Typical droplet excitations dominate the thermodynamic behaviour.
A closely related idea is the domain wall defect \cite{McMillan84x,BM84}; essentially
a droplet perimeter that extends to infinity. With a continuous distribution of disorder these 
related views seem to be equivalent \cite{HY02} according to the predictions of droplet theory.

For the bimodal model domain wall defects have, for example, been applied \cite{KR97,AH04} 
to the determination of the value of the critical defect concentration $p_c$. Nevertheless,
it still remains unclear whether droplet theory is appropriate \cite{Hart08}.
The ground state is not unique and a droplet may represent some different ground state; not
an excitation.

For this study the $L \times L$ square lattice is wound around a cylinder, that is we use periodic 
boundary conditions in one dimension. In the second dimension the system is nested in an
infinite unfrustrated environment. There are no open boundaries. If the circumference $L$
of the cylinder is even then the energy gap is $4J$. Otherwise it is $2J$. 
In the spin glass phase the distribution of degeneracies of the first excited state is extreme
with a long tail representing large values \cite{DTWWTSC04,Wanyok}.
We have also looked at systems with open boundaries and have found extreme distributions
of $2J$ excitations in some agreement with Wang \cite{WangWorm}.

The issue of the size of the energy gap of the bimodal Ising spin glass
dates back to the proposal of Wang and Swendsen
\cite{WS88} that it should be $2J$ in the thermodynamic limit. 
It now seems clear that in fact there is no energy gap at all and the low temperature
specific heat varies as a power law $c_v \sim T^{-\alpha}$. 
The first indications of this appeared in Ref. \onlinecite{Jorg06} 
and confirmation \cite{THM11} from the evaluation of very large Pfaffians has 
recently appeared.

The issue that remains unclear is the value of the critical exponent $\alpha$. For the case
of continuous (Gaussian) disorder, direct calculations 
\cite{Cheung83,HH04} report that the specific heat is linear with $\alpha=-1.0$.
For bimodal disorder Monte Carlo work \cite{KLC07} reports that $\alpha=-4.21$ while
droplet theory \cite{THM11} suggests that $\alpha=-3.0$ although the temperature range used
is extremely narrow.
Other Monte Carlo results \cite{Jorg06,KLC07} for the correlation length with the assumption of
hyperscaling gives $\alpha=-7.1$. Universality is hard to prove.

The exponent $\alpha$ is difficult to estimate. One reason that makes this so for the bimodal 
case is that the specific heat is not normally distributed. We have performed some calculations 
with open boundaries and find that the distribution of the specific heat has a tail for low 
temperature and small values of linear sample size $L$. The methods used were direct evaluation
of Pfaffians as well as summing the density of states \cite{SK93}. Although it is reasonable
to believe that the specific heat will be normally distributed in the thermodynamic limit,
it is not clear what value to use from calculations with finite $L$.

It is at least clear now that the low-temperature
specific heat contains contributions from excitations
having a range of energies. This fits well with droplet theory \cite{THM11} where it is
predicted that $\alpha=1-2/\theta_S$ with the fractal dimension of domain walls 
given by $d_f=2\theta_S$.
If $\theta_S = 0.5$ as reported \cite{SK93,LMMS06,THM11} then $\alpha=-3.0$. However,
other work \cite{Roma07,MH07,WJ07,Anuchit} predicts values $d_f > 1$ that imply $\alpha > -3.0$.
It seems unlikely that droplet theory can predict a value in agreement with $\alpha=-4.21$
or $\alpha=-7.1$.

To obtain a simple description of the ferromagnetic phase we can start with the case of low
defect concentration $p$. The defect bonds are widely separated and the ground state is
unique (aside from global inversion). So the degeneracy of the ground state is $M_0=1$.
We can find first excited states by flipping a spin at either end of a defect bond. Thus
the degeneracy of the first excited state is $M_1=4pN$ where the square lattice has $N$ sites
and $2N$ bonds. The value of the density of states $\frac{M_1}{M_0}$ per spin is $4p$.
We can think of clusters of disorder each composed of one negative bond and two frustrated 
plaquettes.

As the concentration $p<p_c$ increases the clusters of disorder grow in size and influence. 
The distribution of the density of states becomes less normal and its peak moves above $4p$. 
Nevertheless, the rigid lattice still percolates and there remains some finite magnetisation.
With a lattice of finite size, wound in one direction, the $4J$ excitations occur in two classes.
Some are derived locally and are not influenced by the boundary condition; just like the simple
case of low concentration. Others are formed by extending all the way around the system.

\begin{figure}[ht]
\includegraphics*[width=8.5cm]{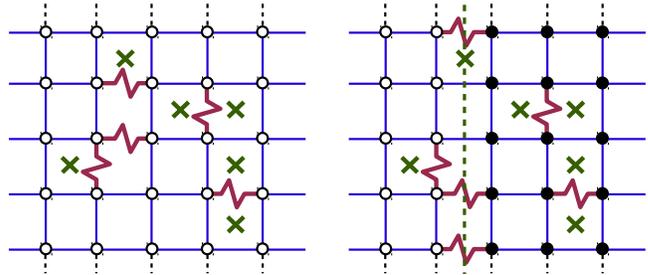}
\caption{\label{f:fig1}(Color online) An example of a $2J$ excitation. On the left is a
ground state configuration with six frustrated plaquettes and five unsatisfied (jagged) bonds.
On the right is a first excited state obtained by flipping all spins on one side of the
vertical broken line. The excited state has six unsatisfied bonds. Periodic BCs are indicated by the top and bottom dashed vertical lines.} 
\end{figure}

Since it is not easy to distinguish between these two classes, we employ the device of fixing
an odd value of the circumference $L$ of the cylinder. In this case the $2J$ excitations are 
entirely nonlocal. Fig. \ref{f:fig1} shows an example. The excitation depends on the boundary 
condition and would not exist otherwise. All $2J$ excitations involve flipping all spins on one side of some closed path around the system. Other closed paths can give excitations with energies equal to an odd multiple of $2J$. A $4J$ excitation requires two paths.

Our main message here is that it is possible to essentially define the ferromagnetic phase by the 
absence of these $2J$ excitations. Alternative approaches \cite{AH04} include the imposition
of domain wall defects and the calculation of magnetisation. These nevertheless lack clear
systematics due to the large degeneracy of the ground state. Domain wall defects may not
represent excitations at all since they can correspond to alternative ground states. 
Sampling of domain walls cannot be done in a controlled way and
it is not obvious \cite{MH07} how we can obtain typical representative domain walls.

Calculation of the magnetisation is also problematic as a result of the ground state 
degeneracy. In Ref. \onlinecite{AH04}, for example, the algorithm starts with a ground
state and proceeds with a Monte Carlo simulation to determine a typical value of the 
magnetisation.

In this work we propose a simple picture of the ferromagnetic phase that is evaluated
from the response to temperature alone. The number of lowest energy 
excitations is counted exactly. In the thermodynamic limit these excitations can only
exist in the spin glass phase. Details of our results are given in Sec. \ref{sec:results} after
a brief account of our method.

\section{Formalism}
\begin{figure}[htb]
  \begin{center}
    \includegraphics[width=3cm]{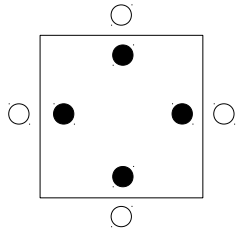}
  \end{center}
  \vspace{-4mm}
  \caption{A plaquette with associated lattice fermions. The filled circles are associated with the plaquette, and the pairs of filled and open circles are associated with the bonds.}
  \label{fig2}
\end{figure}

We use the Pfaffian method and degenerate state perturbation theory to calculate the degeneracies of the excited states. The planar Ising model can be mapped onto a system of noninteracting fermions. Each bond is decorated with two fermions, one either side. A square plaquette then has four fermions inside and four others across the bonds, as shown in Fig. \ref{fig2}. For a system with $N$ lattice sites we have $4N$ fermions in total. The partition function can be written as \cite{GH64,B82}
\begin{equation}
 Z = 2^N \left( \prod _{\langle ij \rangle} \cosh (J_{ij} /kT) \right) (\det D)^{1/2} .
\end{equation}
The product is over all nearest-neighbor bonds $J_{ij}$ on an $N$ site lattice. The matrix $D$ is a $4N\times4N$ skew-symmetric matrix that comprises constant diagonal blocks, and off-diagonal blocks that depend on temperature $T$ through matrix elements $\pm \tanh J_{ij}/kT$. The factor $(\det D)^{1/2}$ is precisely the Pfaffian \cite{GH64,B82}. This formalism is applicable to any distribution of disorder. For the bimodal model $J_{ij} = \pm J$.

At zero temperature there are defect eigenstates of $D$ with eigenvalues equal to zero. Each defect eigenstate can be expressed as a linear combination of the fermions localized in a frustrated plaquette. The number of these defect eigenstates is exactly equal to the number of frustrated plaquettes. At low temperature each defect eigenvalue approches zero as
\begin{equation}\label{eigen}
 \epsilon = \pm \frac{1}{2} X\exp \left( -\frac{2Jr}{kT} \right)
\end{equation}
where $r$ is an integer and $X$ is a real number. These quantities $r$ and $X$ can be obtained using degenerate state perturbation theory \cite{BP91}. The ground state energy is written as
\begin{equation}
  U_0 = -2NJ + 2J \sum_d r_d
\end{equation}
where the sum are over all defect eigenstate pairs. The ground state degeneracy is
\begin{equation}
  M_0 = \prod_d  X_d
\end{equation}
and the ground state entropy can then be written as $S_0 = k\sum_d \ln X_d$.

At arbitrary low temperature the internal energy can be expanded as \cite{Wanyok}
\begin{equation}
 U = U_0 + \sum_{m=1} ^\infty e^{-2Jm/kT} U_m 
\end{equation}
where the coefficient $U_m$ is expressed as
\begin{equation}
 U_m = -2^m J \ \text{Tr} \ R^m \label{uexpand}
\end{equation}
with
\begin{equation}
 R = D_1g_{c1}(1 + D_1G_1)(1 + D_2G_2)\cdots(1 + D_{r_{max}}G_{r_{max}}).
\end{equation}
The $2 \times 2$ block diagonal matrix $D_1$ is defined according to $D = D_0 + \delta D_1$ where $D_0$ is the matrix $D$ when $T = 0$ and $\delta = 1 - \tanh J/kT$. $D_1$ has non-zero matrix elements joining two fermions across bonds only. The $4\times4$ block diagonal matrix $g_{c1}$ is derived from the continuum Green's function \cite{BP91} and has matrix elements connecting the fermions within a plaquette. $D_2$ is given by $D_2 = D_1 g_{c1} D_1$ and, for $r > 2$, $D_r = D_{r-1}(1 + G_{r-2}D_{r-2})\cdots(1 + G_1D_1)g_{c1}D_1$. The Green's function $G_r$ is given by \cite{BP91} 
\begin{equation}
 G_r = -\sum_{i=1}^{N(r)} \lvert r,i\rangle \left( \frac{1}{\epsilon^i_r} \right) \langle r,i\rvert,
\end{equation}
where $\lvert r,i\rangle$ is the ground state defect eigenstate of $D$ with eigenvalue $\epsilon^i_r$. The integer $r$ represents the order of perturbation theory at which the degeneracy is lifted. It is also the index $r$ in Eq. (\ref{eigen}). The total number of such eigenstates is $N(r)$.

The coefficient $U_m$ can also be expressed in terms of the degeneracies of the excited states. We denote the degeneracy of the {\it i}th excited state as $M_i$. The partition function of the bimodal Ising model can be expressed in terms of the degeneracies as
\begin{equation}
 Z = 2M_0 e^{-\frac{U_0}{kT}} \left( 1 + \frac{M_1}{M_0}e^{-\frac{2J}{kT}} + \frac{M_2}{M_0}e^{-\frac{4J}{kT}} + \ldots \right).
\end{equation}
Using some thermodynamic relations together with the expansion of $\ln Z$ using the Taylor series $\ln (1+x) = x - \frac{x^2}{2} + \frac{x^3}{3} - \ldots$, we obtain, for example,
\begin{eqnarray}
 U_1 & = & 2J\left( \frac{M_1}{M_0} \right) \nonumber \\
 U_2 & = & 4J \left( \frac{M_2}{M_0} - \frac{1}{2} \left( \frac{M_1}{M_0} \right)^2 \right) \nonumber \\
 U_3 & = & 6J \left( \frac{M_3}{M_0} - \frac{M_2}{M_0}\frac{M_1}{M_0} + \frac{1}{3} \left( \frac{M_1}{M_0} \right)^3 \right) \label{degen}
\end{eqnarray}
From these relations the ratios $\frac{M_i}{M_0}$ for all excited states can be obtained recursively. Note that $M_1$ is the number of $2J$ excitations.

\section{\label{sec:results}results}
\begin{figure}[ht]
\includegraphics[angle=-90,width=8.5cm]{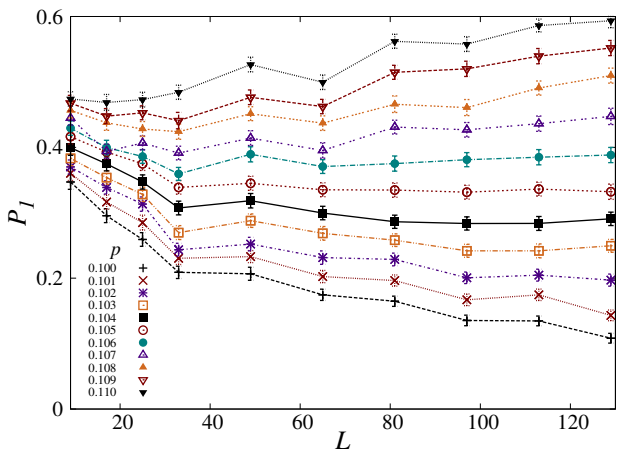}
\caption{\label{f:fig3.ps}(Color online) The probability $P_1$ of finding $\frac{M_1}{M_0} > 0$, plotted as a function of system size $L$ for various values of antiferromagnetic bond concentration $p$.}
\end{figure}

We have calculated $\frac{M_1}{M_0}$ for system sizes up to $L = 129$ and concentrations $p$ ranging from 0.050 to 0.150. The number of disorder realizations ranges from 20000 for the smallest size to 2000 for the largest. We denote as $P_1$ the probability of finding $\frac{M_1}{M_0} > 0$. In Fig. \ref{f:fig3.ps}, $P_1$ is plotted as a function of system size $L$ for various defect concentrations. The error bars are evaluated using the bootstrap method \cite{E82}. The transition concentration $p_c$ is indicated where the $L$ dependency of $P_1$ changes from decreasing to increasing. We can see that $P_1$ is decreasing for $p < 0.102$. The system can be regarded as ferromagnetic below this concentration. Since $P_1$ is increasing with $L$ for $p > 0.106$, the system can be regarded as a spin glass. We conclude from these results that the value of $p_c$ lies between 0.102 and 0.106.

\begin{figure}[ht]
\includegraphics[angle=-90,width=8.5cm]{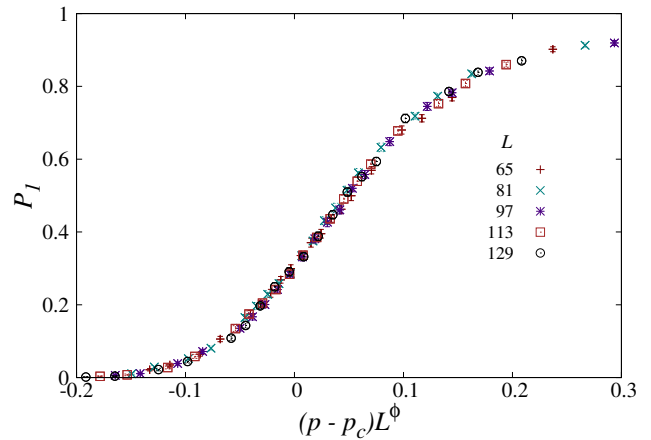}
\caption{\label{f:fig4}(Color online) The scaling plot of $P_1$ as a function of the antiferromagnetic bond concentration $p$ with $p_c = 0.1045(11)$ and $\phi = 0.532(72)$.}
\end{figure}

\begin{figure}[ht]
\includegraphics[angle=-90,width=8.5cm]{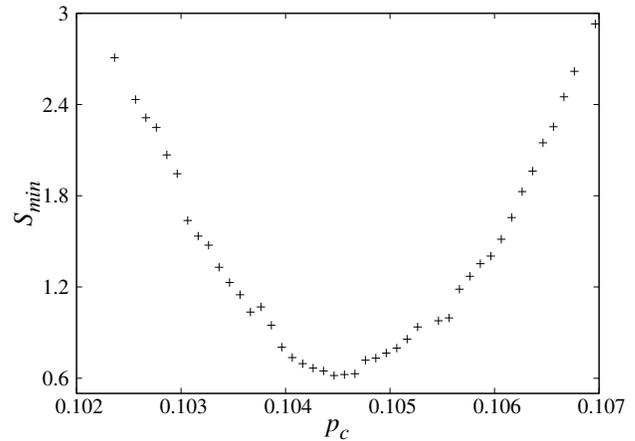}
\caption{\label{f:fig5} The variation of $S_{min}$ as a function of $p_c$.}
\end{figure}

We have done a scaling plot using the relation \cite{KR97},
\begin{equation}
 P_1L^{\psi} = f((p-p_c)L^{\phi}).
\end{equation}
It is reasonable to fix $\psi = 0$ since the value of $P_1$ is bounded to the range [0,1]. In any case, with $\psi$ not fixed, the best scaling plots have $\psi < 0.001$. The parameters $p_c$ and $\phi$ are chosen to minimize the quality parameter \cite{HH04,M09} $S$. The best fits give $p_c = 0.1045(11)$ and $\phi = 0.532(72)$ with $S = 0.62$. The resulting scaling plot is shown in Fig. \ref{f:fig4}. The error bars of each parameter are obtained using the method described in Ref. \onlinecite{AH04}. We fix the corresponding parameter at various values and minimize $S$ with respect to the other parameter. The range of the fixed parameter that gives $S$ double the minimum value is regarded as the error bar. For example we show in Fig. \ref{f:fig5} the variation of the partial minimized value of $S(p_c,\phi)$ as a function of $p_c$. The error bar of $\phi$ can be obtained in the same way.

The above value of $p_c$ agrees, within error bars, with $p_c = 0.103(1)$ proposed in Ref. \onlinecite{AH04}. Note that we also performed the analysis using data from systems with $L \le 65$ and get $p_c \gtrsim 0.105$. This indicates that the effect of finite size is the overestimation of $p_c$. It is expected that if we perform this analysis using data with $L > 129$, we will get a smaller value of $p_c$. %

\begin{figure}[ht]
\includegraphics[angle=-90,width=8.5cm]{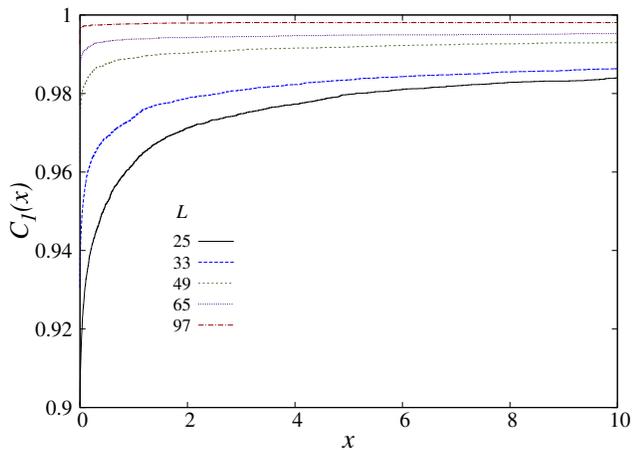}
\caption{\label{f:fig6.ps}(Color online) The probability $C_1(x)$ of finding $\frac{M_1}{M_0} \le x$ for $p = 0.090$.}
\end{figure}

We have also investigated the distributions of the $2J$ excitations in the ferromagnetic phase. We denote as $C_1(x)$ the probability of finding $\frac{M_1}{M_0} \le x$. In Fig. \ref{f:fig6.ps}, $C_1(x)$ with $p = 0.090$ is plotted for various values of $L$. It is clear that the most likely value of $\frac{M_1}{M_0}$ is zero. The probability of getting $\frac{M_1}{M_0} > 0$ is decreasing with $L$. We may expect that in the ferromagnetic phase the $2J$ excitations vanish in the thermodynamic limit. 

\begin{figure}[ht]
\includegraphics[angle=-90,width=8.5cm]{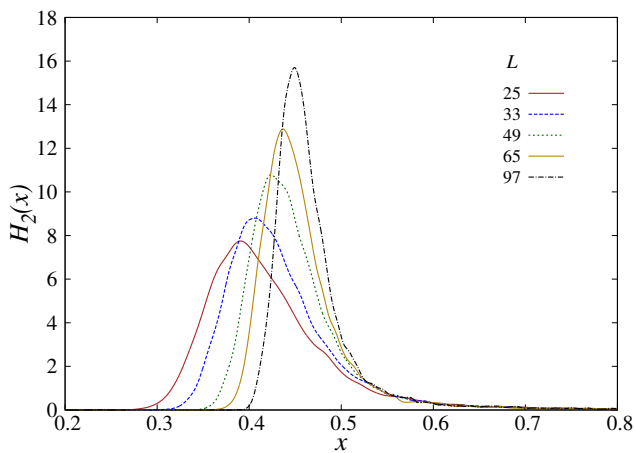}
\caption{\label{f:fig7}(Color online) The probability density $H_2(x)$ of getting $\frac{1}{L^2}\frac{M_2}{M_0} = x$ for $p = 0.090$ with odd $L$.}
\end{figure}

\begin{figure}[ht]
\includegraphics[angle=-90,width=8.5cm]{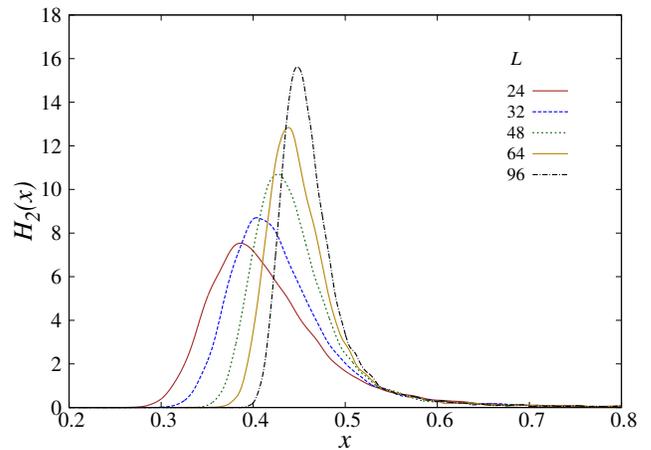}
\caption{\label{f:fig8.ps}(Color online) The probability density $H_2(x)$ of getting $\frac{1}{L^2}\frac{M_2}{M_0} = x$ for $p = 0.090$ with even $L$.}
\end{figure}

Since there are no $2J$ excitations in the ferromagnetic phase in the thermodynamic limit, the first excited state has energy $4J$. We have investigated the behavior of the $4J$ excitations by calculating the ratio $\frac{M_2}{M_0}$ for system sizes up to $L = 97$. We denote as $H_2(x)$ the probability density function of getting $\frac{1}{L^2}\frac{M_2}{M_0} = x$. We use the kernel density estimation algorithm \cite{BGK10} to obtain $H_2(x)$. In Fig. \ref{f:fig7}, $H_2(x)$ with $p = 0.090$ is plotted for various odd values of $L$. A sharp peak develops with increasing $L$. We expect to get a definite value of $\frac{1}{L^2}\frac{M_2}{M_0}$ in the thermodynamic limit. It is interesting that this behaviour does not depend on whether $L$ is odd or even. In Fig. \ref{f:fig8.ps}, $H_2(x)$ with $p = 0.090$ is plotted for various even values of $L$. The distributions are much the same and provide the same conclusions. From these results we have that the energy gap in the ferromagnetic phase is $4J$. 

We can expect this also from the behavior of the specific heat at low temperature. When the temperature is low enough the behavior is dominated by the first excited state and can be expressed as \cite{KLC07}
\begin{equation}
 c_v = \frac{16J^2}{kT^2} \left(\frac{1}{L^2}\frac{M_2}{M_0} \right)e^{-4J/kT}.
\end{equation}
Sharpness of the distribution of $\frac{1}{L^2}\frac{M_2}{M_0}$ satisfies the requirement of $c_v$ as a physical quantity. We can expect that in the ferromagnetic phase the specific heat has a definite value in the thermodynamic limit. At low temperature $c_v$ is proportional to $\exp(-4J/kT)$ and the energy gap can be regarded as $4J$.

\begin{figure}[ht]
\includegraphics[angle=-90,width=8.5cm]{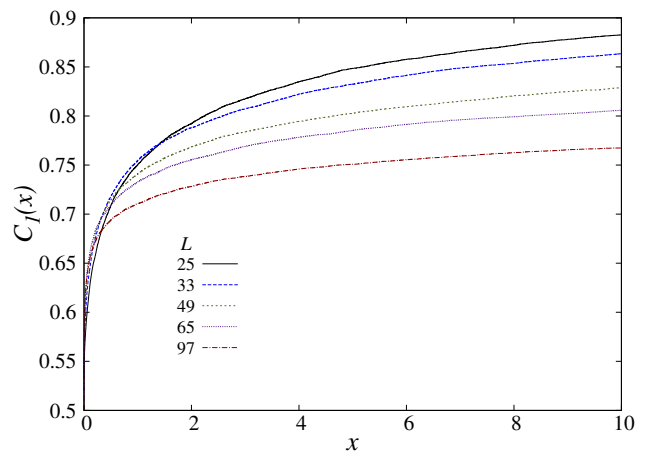}
\caption{\label{f:fig9.ps}(Color online) The probability $C_1(x)$ of finding $\frac{M_1}{M_0} \le x$ for $p = 0.110$.} 
\end{figure}

The distribution of the $2J$ excitations in the spin glass phase is quite different. In Fig. \ref{f:fig9.ps}, $C_1(x)$ with $p = 0.110$ is plotted for various values of $L$. Although the most likely value of $\frac{M_1}{M_0}$ is still at zero, the probability of getting $\frac{M_1}{M_0} > 0$ is increasing with $L$. The distributions of $\frac{M_1}{M_0}$ do not have a sharp peak but broaden when $L$ is increasing. We have that the $2J$ excitations persist as $L$ increases.

\begin{figure}[ht]
\includegraphics[angle=-90,width=8.5cm]{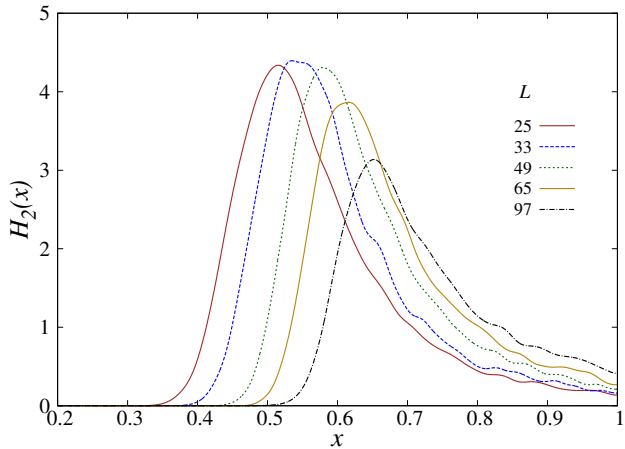}
\caption{\label{f:fig10}(Color online) The probability density $H_2(x)$ of getting $\frac{1}{L^2}\frac{M_2}{M_0} = x$ for $p = 0.110$ with odd $L$.}
\end{figure}

\begin{figure}[ht]
\includegraphics[angle=-90,width=8.5cm]{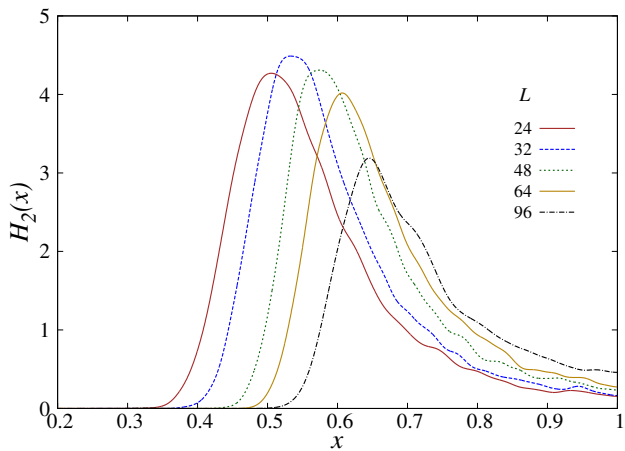}
\caption{\label{f:fig11}(Color online) The probability density $H_2(x)$ of getting $\frac{1}{L^2}\frac{M_2}{M_0} = x$ for $p = 0.110$ with even $L$.}
\end{figure}

The distribution of the $4J$ excitations in the spin glass phase is also different from that in the ferromagnetic phase. In Fig. \ref{f:fig10}, $H_2(x)$ with $p = 0.110$ is plotted for various odd values of $L$. The most likely value of $H_2(x)$ increases with $L$ and the distributions broaden. This behavior of $H_2(x)$ is similar \cite{Wanyok} to that of the canonical spin glass ($p = 0.5$) with even $L$. In particular, the height of the peak of the distribution collapses with increasing $L$. We have also checked the distributions of $H_2(x)$ with $p = 0.110$ and even $L$. The results are shown in Fig. \ref{f:fig11}. The characteristics are the same for both odd and even $L$. 

\section{Summary}
We have proposed a simple view that distinguishes between the ferromagnetic and spin glass
phases. The ferromagnetic phase is characterised by the absence of lowest energy,
that is $2J$ excitations. Our method counts the number
of excitations exactly without bias. It is not necessary to work with some typical ground 
or excited state.

Distributions of the number of $2J$ excitations are shown to differ in character between
the phases. In the ferromagnetic phase the number declines as the (odd) circumference $L$ of the
cylindrical winding increases. A finite-size scaling analysis produces a data collapse of
excellent quality to support our conclusion that $2J$ excitations do not exist in the 
thermodynamic limit of the ferromagnetic phase. In the spin glass phase the situation is 
reversed with the degeneracy of the first excited state increasing with $L$. 

The energy gap in the ferromagnetic phase is $4J$. For even values of $L$ the first
excitations have energy $4J$. We have also presented distributions of $4J$ excitations
so as to indicate that there is no essential dependence on whether $L$ is even or odd.
In the ferromagnetic phase the peak grows taller and narrower with increasing $L$ and
will presumably lead to a unique value of the low temperature specific heat in the
thermodynamic limit. 

In the spin glass phase the behaviour of the distributions is quite different. Essentially they
are extreme with long tails. As $L$ increases the tails become fatter and the peak collapses.
We believe that this is consistent with a power law behaviour for the low temperature
specific heat. The extreme distributions only indicate spin glass behaviour; a proper
statistical mechanical description of the model requires a summation over the entire
density of states. This seems to suggest that thermally active droplets can indeed take many 
different values of energy.

Finally, we have not found any evidence that indicates a random antiphase state \cite{BMR82}, although we cannot rule out a situation where percolation of rigid bonds coexists with zero magnetization.

\begin{acknowledgments}
N. J. thanks the National Science and Technology Development Agency, Thailand for a scholarship.
Some of the computations were performed on the Tera Cluster at the Thai National Grid Center and on the Rocks Cluster at the Department of Physics, Kasetsart University.
\end{acknowledgments}

\end{document}